\documentstyle[11pt,newpasp,twoside,epsf]{article}
\def\emphasize#1{{\sl#1\/}}
\def\arg#1{{\it#1\/}}
\let\prog=\arg

\def\edcomment#1{\iffalse\marginpar{\raggedright\sl#1\/}\else\relax\fi}
\reversemarginpar

\begin{document}
\title{The Nature of Flat-Spectrum Nuclear Radio Emission in Seyfert Galaxies}
 \author{C.G. Mundell$^{1,2}$, A.S. Wilson$^2$, J.S. Ulvestad$^3$ \& A.L. Roy$^{3,4}$}

\affil{$^1$ARI, Liverpool John Moores University, U.K. (cgm@astro.livjm.ac.uk)}
\affil{$^2$University of Maryland, College Park, MD20742}
\affil{$^3$NRAO, P.O. Box O, Socorro, NM87801}
\affil{$^4$MPIfR, Bonn, Germany}

\begin{abstract} Parsec-scale VLBA imaging of five Seyfert galaxies
with flat-spectrum radio nuclei was conducted to determine whether the
flat spectrum represents thermal emission from the accretion
disk/obscuring torus or nonthermal, synchrotron self-absorbed
emission. Four of the five show emission consistent with synchrotron
self-absorption, with intrinsic sizes $\sim$0.05$-$0.2 pc (or 10$^4$
gravitational radii for a 10$^8$ M$_\odot$ black hole for the
smallest). In contrast, NGC~4388, which was detected with MERLIN but
not the VLBA, shows thermal emission with similar properties to that
detected in NGC~1068. It is notable that the two Seyfert galaxies with
detected thermal nuclear radio emission both have large X-ray
absorbing columns, suggesting that columns in excess of
$\sim$10$^{24}$~cm$^{-2}$ are needed for such disks to be detectable.
\end{abstract}

\section{Introduction}

AGN are thought to be powered by accretion of material onto a central
supermassive black hole via a disk that regulates the fueling rate.
The extent of these disks is not well established but AGN unification
schemes advocate a geometrically thick or warped thin disk that hides
the nucleus when viewed edge-on and accounts for the observed
differences between broad (type 1) and narrow-line (type 2) AGN.
Theoretical work indicates that UV/X-ray radiation from the central
engine can heat, ionize and evaporate gas on the inner edge of the
torus$^{1}$ and recent high angular resolution VLBA radio observations
of the archetypal Seyfert 2 galaxy, NGC~1068, have shown that emission
from the flat-spectrum radio component `S1' may be associated with
the thermal gas on the inner, ionized edge of the torus$^2$.  This
discovery highlights the possibility of using the VLBA to image the
pc-scale disks or tori in other Seyfert galaxies.  Flat-spectrum
nuclei in radio galaxies and quasars often represent non-thermal
synchrotron self-absorbed radio emission with a much higher brightness
temperature ($>$10$^8$ K) than is characteristic of component S1 in
NGC 1068, so high resolution radio observations are required to
distinguish between the two emission processes.

\section{The Search for Thermally-Emitting Disks$^3$}

Four of five sources imaged with the VLBA at 8.4 GHz were detected
(Figure 1) and show compact, unresolved nuclei with brightness
temperatures, $T_{\rm B}$~$>$~10$^8$ K, monochromatic luminosities
$\sim$10$^{21}$ W Hz$^{-1}$ and sizes less than $\sim$1~pc. We
conclude that the sub-pc scale radio emission in these sources is
non-thermal and self absorbed, and hence dominated by the central
engine.  We find no significant evidence of thermal disk-like emission
extended perpendicular to the collimation axis in these sources.  In
contrast, the nucleus of NGC~4388 is not detected with the VLBA but is
detected with MERLIN at 5~GHz. The inferred brightness temperature of
2.4~$\times$~10$^4$~K~$<$~$T_{\rm B}$~$<$~2.2~$\times$~10$^6$~K is too
low for synchrotron self absorption to be important and we propose a
model in which the emission is optically thin thermal bremsstrahlung
from a gas with electron temperature of $T_{\rm e}$~$>$~10$^{4.5}$~K
and density $n_e$~$>$~1.6~$\times$~10$^4$~f~$^{-0.5}$~cm$^{-3}$ (f is
the volume filling factor). The larger inferred values of $T_{\rm
e}$~$=$~10$^{6.8}$~K and
$n_e$~$=$~1.8~$\times$~10$^6$~f~$^{-0.5}$~cm$^{-3}$ for the smaller
source size set by the VLBA limit, are similar to the values of
$\sim$10$^{6.8}$~K and $\sim$10$^{6.8}$~cm$^{-3}$ found for thermal
emission in NGC~1068$^2$, thus implying that we may be seeing the same
phenomenon in NGC~4388.  It is notable that these two Seyferts with
detected thermal nuclear radio emission have large X-ray absorbing
columns, suggesting that columns in excess of $\sim$10$^{24}$~cm$^{-2}$
are needed for such disks to be detectable.

\begin{figure}  
\plotfiddle{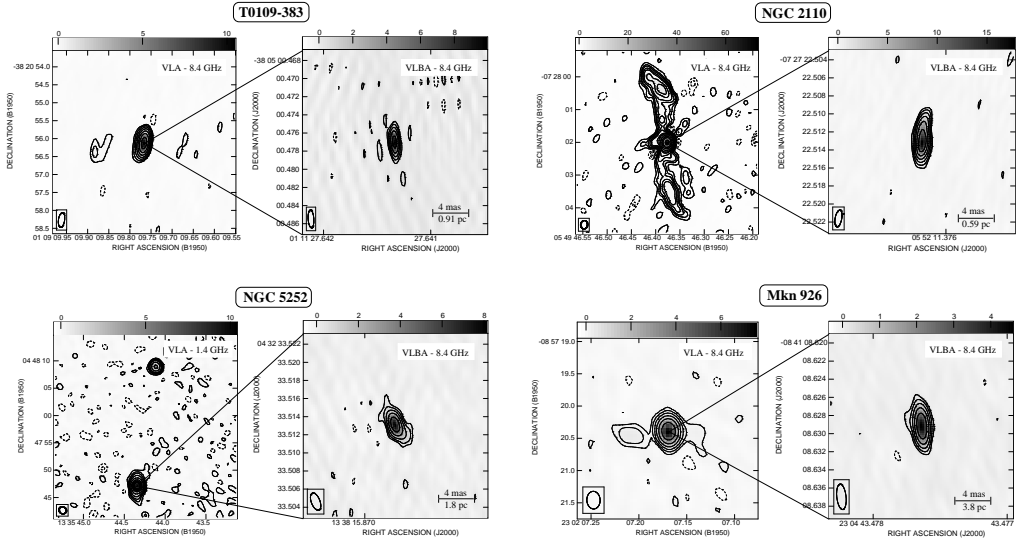}{6.2cm}{270}{53}{53}{-215}{255}
\caption{\small 8.4-GHz continuum images (right-hand panel) of sources
detected with VLBA: T0109-383, NGC 2110, NGC 5252 and Mkn 926; VLA
image of larger scale structure also shown (left-hand panel).}
\end{figure}

\end{document}